\documentclass[sigconf,authorversion]{acmart}

\usepackage{algorithmic}
\usepackage{graphicx}
\usepackage{textcomp}
\usepackage{xcolor}

\usepackage{listings}
\usepackage{verbatim}
\usepackage{booktabs}
\usepackage{tabularx}
\usepackage{booktabs}
\usepackage{color}
\usepackage{upquote}
\usepackage{balance}
\usepackage{hyperref}
\usepackage{xspace}
\usepackage[ruled,linesnumbered]{algorithm2e}
\usepackage{caption}
\usepackage{setspace}

\newcommand{\basenamerandomized}{Randomized}
\newcommand{\randomized}{\basenamerandomized\xspace}

\newcommand{\basenamefixed}{Fixed}
\newcommand{\fixed}{\basenamefixed\xspace}

\copyrightyear{2024}
\acmYear{2024}
\setcopyright{rightsretained}
\acmConference[RAID 2024]{The 27th International Symposium on Research in Attacks, Intrusions and Defenses}{September 30-October 02, 2024}{Padua, Italy}
\acmBooktitle{The 27th International Symposium on Research in Attacks, Intrusions and Defenses (RAID 2024), September 30-October 02, 2024, Padua, Italy}
\acmPrice{}
\acmDOI{10.1145/3678890.3678931}
\acmISBN{979-8-4007-0959-3/24/09}

\AtBeginDocument{
  \providecommand\BibTeX{{
    \normalfont B\kern-0.5em{\scshape i\kern-0.25em b}\kern-0.8em\TeX}}}

\begin{document}

\title{Hidden Web Caches Discovery}

\author{Matteo Golinelli}
\email{matteo.golinelli@unitn.it}
\orcid{0000-0002-8743-0825}
\affiliation{
  \institution{University of Trento}
  \city{Trento}
  \country{Italy}
}

\author{Bruno Crispo}
\email{bruno.crispo@unitn.it}
\orcid{0000-0002-1252-8465}
\affiliation{
  \institution{University of Trento}
  \city{Trento}
  \country{Italy}
}

\renewcommand{\shortauthors}{Golinelli et al.}

\begin{abstract}
Web caches play a crucial role in web performance and scalability. However, detecting cached responses is challenging when web servers do not reliably communicate the cache status through standardized headers. This paper presents a novel methodology for cache detection using timing analysis. Our approach eliminates the dependency on cache status headers, making it applicable to any web server. The methodology relies on sending paired requests using HTTP multiplexing functionality and makes heavy use of cache-busting to control the origin of the responses.
By measuring the time it takes to receive responses from paired requests, we can determine if a response is cached or not. In each pair, one request is cache-busted to force retrieval from the origin server, while the other request is not and might be served from the cache, if present. A faster response time for the non-cache-busted request compared to the cache-busted one suggests the first one is coming from the cache. We implemented this approach in a tool and achieved an estimated accuracy of 89.6\% compared to state-of-the-art methods based on cache status headers. Leveraging our cache detection approach, we conducted a large-scale experiment on the Tranco Top 50k websites. We identified a significant presence of hidden caches (5.8\%) that do not advertise themselves through headers. Additionally, we employed our methodology to detect Web Cache Deception (WCD) vulnerabilities in these hidden caches. We discovered that 1.020 of them are susceptible to WCD vulnerabilities, potentially leaking sensitive data. Our findings demonstrate the effectiveness of our timing analysis methodology for cache discovery and highlight the importance of a tool that does not rely on cache-communicated cache status headers.
\end{abstract}



\settopmatter{printfolios=true}

\maketitle

\section{Introduction}
\label{sec:introduction}

Web caches are servers placed between a client and an origin server that store copies of responses to improve the websites' performance, availability and scalability. Moreover, web caches reduce the load imposed on origin servers by directly serving resources that they previously cached to the clients. For these reasons, web caches have become crucial components of modern web architectures, and are ubiquitous. Web caches can be placed anywhere on the path between the client and the origin server, and frequently multiple caches coexist in the same client-server path.
To communicate whether a response is coming from the origin server (a cache \texttt{MISS}), or if it was cached (a cache \texttt{HIT}), caches employ specific response headers called \textit{cache status headers}. These headers are not standardized, therefore, different cache technologies might use different and custom header names and values. Previous studies have developed heuristics to analyse these headers and distinguish between cached and non-cached responses~\cite{mirheidari2022web}.
Web caches employ unique identifiers called \textit{cache keys} to identify cached resources. They comprise the elements of HTTP requests that must match in different requests for the web cache to issue the same response. They generally include the path, the query string, and the value of specific headers. The elements of a request that are included in the cache key are called \textit{keyed}. To forcefully receive a fresh response from the origin server even if a web cache already holds a stored copy, we can use \textit{cache-busting}, a technique that consists of including random modifications to specific keyed elements of requests (without introducing modifications that cause the response content to be different compared to the non-cache-busted request). Cache busting is useful when we need to test for web cache behaviours.

Techniques that detect cached responses based on cache status headers are not effective when these headers are missing, wrong, or use custom names and values that are not covered by the heuristics used. For this reason, in this research, we develop a novel methodology that uses timing analysis to distinguish between cached and non-cached responses, that can work against any web server, regardless of whether it communicates the cache status of responses or not. Our methodology is based on repeatedly sending paired requests to a web server. When the two requests reach a web cache at the same time and are processed concurrently, if only one of the two is cached, it will consistently return to the client first and faster. If, instead, both responses are coming from the origin server, they will arrive at the client with an inconsistent order and timing.

To control which responses will be served by the origin server and which by the web cache, we make heavy use of cache-busting techniques. Before developing our methodology, we carried out a preliminary experiment of different cache-busting techniques to identify the most effective against a higher number of websites in the wild. Combining all the techniques that we used, we are able to cache-bust requests on 84.3\% of websites. Our methodology relies on the multiplexing functionality to pair requests together and send them in a single packet. We focus on HTTP/2 since it is the most adopted version, but the same methodology applies to HTTP/3 too.

Our methodology works as follows: we send two groups of \texttt{n} paired requests. In the first group, all requests are cache-busted (i.e., all responses come from the origin server), while for each pair in the second group, only one request is cached-busted (i.e., one is possibly coming from a web cache). We call the two groups \randomized and \fixed, respectively. The idea is that, if the non-cache-busted request is cached, the timing measurements in the two groups are significantly different. To check that, we use a statistical test (\textit{t-test}).

We implement this methodology in a tool and perform a preliminary experiment to measure its accuracy compared with the state-of-the-art heuristics based on cache status headers, estimating an accuracy of 89.6\%. However, we observed that in a vast majority of the cases where our methodology classified the request as cached, while the cache status headers reported a cache MISS, it was due to an unpredictable behaviour of certain web cache technologies. These caches always report cache MISS when two requests are paired together, even if one of the two responses was a cache HIT. We randomly selected 100 websites and manually verified that in 82 of them, the wrong classification of our methodology was due to the behaviour mentioned above. We can therefore estimate that the real accuracy of our tool is higher, and the reported accuracy should be considered as a lower bound.

We then used our methodology to estimate the prevalence of \textit{hidden web caches} in the Tranco Top 50k, i.e., caches that do not advertise the status of their responses in the cache status headers, finding that 1.627 websites (5.8\% of the 28.243 tested websites that supported HTTP/2) present a hidden cache.

Finally, we use our novel timing analysis to detect caching to create a methodology that can detect Web Cache Deception vulnerabilities in a black-box manner, and use it to test these 1.627 previously identified hidden caches.
We find that 1.020 of them cache dynamic content, that they should not be cached, and we present case studies of Web Cache Deception vulnerabilities that we successfully exploited to leak sensitive data of our test victim accounts.
These WCD vulnerabilities could not have been detected with the previous state-of-the-art methodologies, highlighting the importance of a methodology that does not rely on the cache status headers.

\subsection{Contributions}

To summarize, we make the following contributions:

\begin{itemize}
    \item We present a novel methodology to detect caching using timing analysis. Our methodology is simple, does not rely on cache-communicated cache status headers, and applies to all web servers that use newer versions of HTTP.
    \item We conduct an experiment on the effectiveness of different cache-busting techniques, to identify the most commonly keyed elements of HTTP requests. We find that modifying the query string is the most effective and that, combining all techniques, we can cache-bust requests on 84.3\% of the tested websites.
    \item We conduct a large-scale experiment on the Tranco Top 50k using our novel detection methodology to measure the prevalence of hidden caches, i.e., web caches that do not provide the cache status headers of their responses. We detect hidden caches on 1.627 websites.
    \item We use our novel cache detection methodology to test websites for Web Cache Deception (WCD) vulnerabilities. We find that 1.020 websites with hidden caches are vulnerable to WCD vulnerabilities. We present case studies of well-hidden vulnerabilities that, without our methodology, could not have been identified.
\end{itemize}

\textbf{Availability} The code used for this research is available as an open-source tool on the authors' websites.~\footnote{https://github.com/golim/hidden-web-caches-discovery}

\section{Background}
\label{sec:background}

This section presents an overview of web caches and Web Cache Deception vulnerabilities, HTTP/2 and timing attacks.

\subsection{Web caches and reverse proxies}

Web caches are intermediary servers that store frequently accessed web content to enhance website performance, lowering loading times, and reducing the computing loads on the origin servers. Web caches can be deployed in multiple stages between the client and the origin server, including the browser. Moreover, a path between a client and a server might present more than one web cache, potentially managed by different entities and organizations.

Web caches effectively act as proxies between clients, such as web browsers, and origin servers, directly serving the responses that they previously cached. When a client requests a web resource, the proxy intercepts the request and checks if it already holds a cached copy of the response. If the content is in the cache and hasn't expired, the cache delivers it directly to the client. Otherwise, the cache issues a request to the origin server. When it receives the response, it forwards it to the client and, if it matches some pre-configured criteria, caches it for future visitors.

\paragraph{Content Delivery Networks}
A Content Delivery Network (CDN) is a geographically distributed network of servers that deliver web resources (e.g., web pages, images, style sheets, script files) to users faster and more reliably than a sole origin server. In CDNs, servers are strategically located at geographically distributed data centres. This distribution places content physically closer to end-users, significantly reducing the distance data needs to travel, and consequently improving website loading times, leading to a better user experience.

\paragraph{Cache Key}
To understand whether the cache already holds a cached response for a request, web caches employ cache keys. A cache key is a unique identifier assigned to a piece of stored data in a web cache. Cache keys include some specific elements of HTTP requests that must match in subsequent requests to issue the same cached response. Cache keys generally include the URL path of the request and the query string, but can also be configured to include specific headers, such as the \texttt{Cookie} and the \texttt{Origin}. The elements of requests included in the cache key are called \textit{keyed}. The process of introducing modifications to the keyed elements of  HTTP requests is called \textit{cache busting}.

\paragraph{Cache status headers}
Cache status headers are used by web caches and CDNs to communicate whether a response is coming from a web cache or the origin server. Cache status headers are not standardized, therefore, each caching technology might use different header names and values. Previous studies have highlighted the cache status headers names and values of the most popular caching technologies and developed heuristics to read them and understand the source of the responses~\cite{mirheidari2022web}.

\subsection{Web Cache Deception}
Web Cache Deception vulnerabilities arise when a web cache and an origin web server disagree on whether a resource is cacheable or not. As a result, an attacker can exploit WCD vulnerabilities to induce a cache into storing content that should be considered private, rendering it publicly accessible.
Static content is the same for all visitors of a website, and cannot include sensitive information by definition; therefore, we can define WCD as the erroneous caching of dynamically generated content. To exploit WCD vulnerabilities, attackers generate an attack URL (generally comprising of a non-existent file name and a static file extension) and use social engineering techniques to induce a victim into visiting it. When the victim visits the attack URL, the origin server generates a response that includes their personal information. If the cache is configured to cache content based on its URL, it will see the static file extension and cache it, making it publicly accessible to the attacker through the same attack URL.

\subsection{HTTP/2}
Hypertext Transfer Protocol (HTTP) is the application-level protocol that is used to transfer data on the web. HTTP/2 was released in 2015 and was the first major update since HTTP/1.1, which was first published in 1997. HTTP/2 is based on TLS over TCP and maintains the semantics of HTTP/1.1, but it changes the way that the data is transferred. While HTTP/1.1 was a plain-text human-readable protocol, HTTP/2 is a binary protocol. HTTP/2 implements optimized mappings of the HTTP/1.1 semantics to enable efficient use of the connection, allowing multiple concurrent requests and responses to be multiplexed over a single TCP connection and compressing the headers. Requests multiplexing consists of the organization of HTTP messages in streams, that are bidirectional sequences of frames with the same identifier, generally representing a request-response pair. Frames are the smallest protocol unit in HTTP/2: they can be of different types (e.g., data, headers, settings) and have an id that identifies the stream to which they belong~\cite{rfc9113}. Web servers process the requests as soon as they have all the frames, and send the response as soon as it is generated.

In 2022, the Internet Engineering Task Force (IETF) published HTTP/3, which differs from HTTP/2 in that it uses the QUIC transport protocol instead of TCP. QUIC is a transport protocol that runs on top of UDP and provides multiplexing, encryption and congestion control directly. HTTP/3 was developed to solve the problems caused by the fact that TCP has no visibility over HTTP/2 multiplexing. Therefore, some features of HTTP/2 are delegated to QUIC in HTTP/3 (i.e., multiplexing and flow control), while others are implemented on top of it~\cite{rfc9114}.

\subsection{Timing Attacks}
Timing attacks focus on indirect leaks of information; specifically, the time it takes a system to perform certain tasks. By measuring the time variations between the execution of different actions, attackers can potentially extract sensitive data from a system. Timing attacks are more effective when performed locally, due to the absence of network jitter and delay, but previous studies during the last twenty years have shown that they are a viable attack vector over the network too. Timing attacks leverage the unintentional side effects of a system's operation. This makes them particularly hard to detect and requires careful design and implementation of security measures in software systems to prevent time-based leaks.

Timeless Timing Attacks (\textit{TTA}), first introduced by Van Goethem et al.~\cite{vangoethem2020tta}, are a novel type of timing attacks that improve the accuracy and greatly lower the required number of requests. They are based on measuring the relative timing difference between requests that are processed concurrently by the web server, while classical timing attacks consist of independent measurements over the network. To make the two executions concurrent, this attack technique sends the two requests in a single packet, enabling attackers to observe all the timing differences greater than the network jitter introduced once the requests arrive at the server, such as the delay introduced by the network card, decryption and ordering of packets. Moreover, timeless timing attacks observe the response packets sequence number (which is monotonically increasing in TCP) to identify the request that the server finished processing first. To send two requests in a single packet, an attacker can exploit HTTP/2 and HTTP/3's multiplexing functionalities. This way, two paired requests will reach the server at the same time and, ideally, be processed concurrently.

\section{Related Works}
\label{sec:related_works}

This section provides an overview of the related works on timing attacks and web cache attacks and vulnerabilities.

\subsection{Timing Attacks}

To develop our timing analysis methodology, we take inspiration from timing attacks.
Timing attacks have been known and used since 1996 when Kocher introduced them and showed how they could be used to find Diffie-Hellman exponents, factor RSA keys, and break other cryptosystems~\cite{kocher1996}. Initially, timing attacks have only been used to break cryptosystems locally~\cite{dhem2000, schindler2000, schindler2002optimized}.
Later, several studies demonstrated the practicality of timing attacks over the network, enabling extracting private keys from network servers~\cite{BRUMLEY2005701, aciicmez2005, bernstein2005cache, brumley2011, crosby2009}.

Felten and Schneider show how to exploit a browser's cache from a malicious web page to determine if a user had recently visited another unrelated web page by issuing requests and checking if the time required to get the response is less than a threshold~\cite{felten2000}. Bortz et al. show two types of timing attacks against websites: direct timing, where private information is leaked directly by the attacker measuring the response time from the server, and cross-site timing, where a malicious website obtains information on a different website from the user's perspective~\cite{bortz2007}. Jia et al. show how to infer the geographical location of victims using timing attacks, exploiting the browser's cache~\cite{jia2015}. Gelernter and Herzberg bypass the same-origin policy and extract sensitive information by measuring the time it takes for the browser to receive the responses to search queries~\cite{gelernter2015}. Van Goethem et al. show that modern browsers expose new side channels that can be used to acquire accurate timing measurements regardless of network contentions and analyse new browser features that can be exploited to obtain substantially more timings~\cite{vangoethem2015}. All major browsers implement defence mechanisms to protect against timing attacks based on lowering the resolution of the timing information; however, Schwarz at al. prove this approach ineffective and present new mechanisms to obtain absolute and relative timings~\cite{schwarz2017}. Smith et al. propose attacks to leak the browsing history of victims~\cite{smith2018}, and Sanchez-Rola et al. show a way to fingerprint hardware devices timing the execution of cryptographic browser API functions that can be mounted remotely with a malicious website and JavaScript code~\cite{Sanchez-Rola2018}. Vanderlinden et al. use the \texttt{server-timing} header value, which generally provides server-side timing information accurate to the millisecond, to reduce the impact of jitter on remote timing attacks. They show that this enables significantly reducing the required number of requests for a successful attack~\cite{vanderlinden2023can}. Similarly, Vanderlinden et al. use timing information exposed in HTTP response headers by backend servers to reduce the jitter included in an attacker's sample. Specifically, they use the timestamp of responses included by web servers in the \texttt{Date} header to synchronize the attacker to the target server, improving classical timing attacks by reducing the number of requests necessary for a successful attack~\cite{vanderlinden2024}.

\subsection{Web Cache Attacks}
Web cache poisoning is an attack that consists of injecting malicious content into a web cache that is then served to unsuspecting victims. Kettle presents a methodology to detect cache poisoning vulnerabilities and shows attacks against popular websites and caching technologies~\cite{kettle2018practical, kettle2020web}. Chen et al. present cache poisoning attacks exploiting inconsistencies in the interpretation of the host header by different HTTP implementations~\cite{chen2016host}. Nguyen et al. show how cache poisoning vulnerabilities can lead to Denial of Service~\cite{nguyen2019your}.

In 2017, Gil introduces Web Cache Deception attacks in~\cite{gil2017wcd}. Mirheidari et al. present an automated detection methodology to detect WCD vulnerabilities and use it to measure their prevalence in a large-scale measurement~\cite{mirheidari2020cached}. In 2022, Mirheidari et al. present a novel methodology which requires no authentication to detect WCD vulnerabilities and show how these vulnerabilities can also be used to steal security tokens from victims, leading to severe consequences for their security. Their methodology is based on the lookup of cache status headers in responses and content identicality checks~\cite{mirheidari2022web}.

\section{Research Goals}
Previous studies have used lookups of the cache status headers to distinguish cached responses from responses coming from the origin server. This task is frequently needed for the detection of several web cache vulnerabilities, such as Web Cache Deception and cache poisoning. However, the algorithms to detect if a request is cached fall short when the cache status headers are missing, or wrong, or use custom names and values for the headers. This might happen for several reasons. First, websites might want to hide the presence of a web cache to avoid attacks, in an attempt to obtain security through obscurity. Moreover, custom web caches and uncommon caching technologies might use uncommon names and values for the cache status headers that were not previously seen by the developers of cache detection techniques. For this reason, we develop a methodology to identify cached responses and distinguish them from non-cached ones without relying on cache status headers. We refer to the web caches that do not advertise the cache status of their responses in the headers as \textit{hidden caches}.

\subsection{Research Questions}

The goal of our research is to answer the following research questions.

\renewcommand{\theenumi}{(Q\arabic{enumi})}
\renewcommand{\labelenumi}{\theenumi}

\begin{enumerate}
    \item \label{goal:one} Can timing analysis be used to detect whether a response is cached or if it is coming from the origin server? How accurate are they?
    \item \label{goal:two} How many websites on the Tranco Top 50k use a web cache but do not communicate it with cache status headers?
    \item \label{goal:three} Are hidden web caches (i.e., the caches that do not issue cache status headers) vulnerable to common cache vulnerabilities?
\end{enumerate}

\renewcommand{\theenumi}{(\arabic{enumi})}
\renewcommand{\labelenumi}{\theenumi}

\section{Methodology}
\label{sec:methodology}

Our novel methodology to detect web caches employs timing analysis and is composed of two main phases. In the first phase, we collect the timing measurements, exploiting HTTP multiplexing and using cache busting. In the second phase, we read the timings collected in the first phase to infer if there is a cache or not.

\subsection{Collection of Timing Measurements}
\label{sec:methodology:phase_1}

\begin{figure*}
    \centering
    \includegraphics[width=1.0\textwidth]{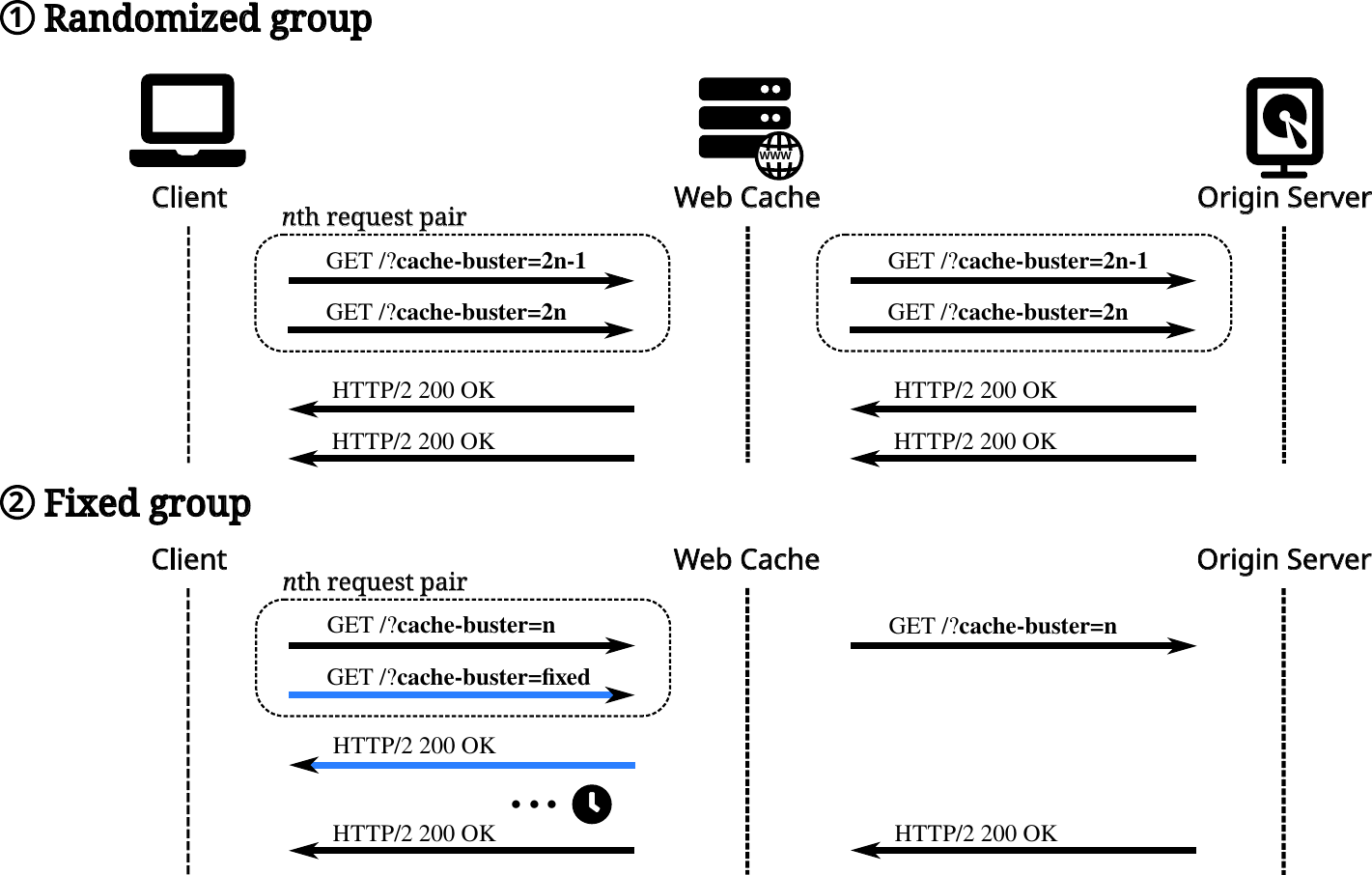}
    \Description[High-level overview of the timing analysis to detect cached responses.]{Longer description.}
    \caption{Overview of our cache detection methodology. Note that, for the \fixed group, we perform a request with fixed cache-busters before collecting the time measurements, so that the response should already be stored in the cache. We see that, in the \randomized group, all requests are forwarded to the origin server, and their order of arrival back at the client is inconsistent. For the \fixed group, instead, the response to the request with a fixed cache-buster is directly issued by the web cache, and will therefore consistently arrive at the client first and faster.}
    \label{fig:tta}
\end{figure*}

In this first phase, we collect the measurements that will be used in the second phase to infer the presence of a cache. This phase is based on HTTP multiplexing and cache busting, respectively used to send the two requests within a single packet and ensure that the responses are served by the origin server and not by the cache. An overview of the timing analysis that we perform is presented in Figure~\ref{fig:tta}.
The idea behind our methodology is to send:

\begin{enumerate}
    \item \textbf{\texttt{n}} pairs of requests in a single packet where both requests have random cache busters (i.e., the responses should always be served by the origin server). We call this group of request pairs \randomized.
    \item \textbf{\texttt{n}} pairs of requests in a single packet where the first request has a random cache buster and the second request has an already used cache buster (i.e., that was included in a previously sent request). In this way, the first response should be served by the origin server, and the second by the cache. We call this group of request pairs \fixed.
\end{enumerate}

In our methodology, \textbf{n} is a fixed number decided before the experiment.
We observe the order of the responses and measure their arrival time difference. By comparing the order of arrival of responses and the time measurements of the two groups, we can determine whether there is a cache in the path from the client to the origin server. Note that we do not measure the absolute time required for responses to arrive but only the relative time elapsed between receiving one response and the other.

\subsubsection*{Cache Busting}

Since we do not have visibility over what fields are included in the cache key by the web caches, we introduce modifications in all the fields that we can modify without obtaining a request for a different resource (e.g., we do not modify the path of the request and the \texttt{Host} header).
Specifically, we introduce modifications in the \textit{query string} by including new parameters with random names and values, and in the following request headers:

\begin{itemize}
    \item \texttt{Origin}: all web caches should include the value of the Origin header in their response so as not to risk introducing Denial of Service vulnerabilities, as shown in~\cite{golinelli2023mind}. Since an Origin is defined as the protocol, the host and the port, we do not modify these to avoid receiving a different response. Instead, we include a randomly generated path in the value of the Origin header, that will not influence the response but is likely included in the cache key.
    \item \texttt{User-Agent}: including the User-Agent in the cache key should not be necessary if the web pages correctly implement and use responsive designs. Since User-Agent values are extremely varied, including them in the cache key might lower the effectiveness of a web cache.
    \item \texttt{X-Forwarded-Host} and \texttt{X-Forwarded-Scheme}: used by proxies to communicate the original host and scheme to the origin server (that might differ). These headers are generally included in the default cache key of caches and CDNs.
    \item \texttt{X-Method-Override}: used by web frameworks to override the HTTP methods of requests. It is generally included in the cache key to avoid cache poisoning vulnerabilities.
\end{itemize}

Moreover, we include random modifications in the headers' values in the \texttt{Vary} response header. The \texttt{Vary} response header is used to communicate what parts of a request may induce differences in the server's responses (not including the method and the URI). Therefore, this header is generally used to communicate to a cache what values of a request must coincide with the ones of the cached response for it to serve the response (i.e., what values should be included in the cache key)~\cite{rfc9110}. For example, if a response contains the \texttt{Vary: Accept-Encoding} header, the cache should only serve the response if the \texttt{Accept-Encoding} header of the request is the same as the one of the cached response.

To evaluate the effectiveness of the different cache-busting techniques, we conducted an experiment on the Tranco Top 10k to identify what elements of HTTP requests are included in their cache key. For each website, we:

\begin{enumerate}
    \item Crawl the website and identify a cached response using \textit{Cache Header Heuristics}, i.e., a lookup of the cache status headers of the responses.
    \item Send one request for each cache-busting technique tested; i.e., we send an HTTP request introducing a modification to a single element of the request.
    \item Check if the response for each request is cached or not. If it is not cached, the modified element is included in the cache key and the cache-busting technique works, otherwise, the element is not part of the cache key.
\end{enumerate}

\begin{table}[t]
    \caption{Results of our experiment on different cache-busting techniques on the Tranco Top 10k. Percentages are calculated over the total number of  3494 websites. Note that only 3128 websites included the \texttt{Vary} header in their responses. Collectively, with the techniques that we employed, we were able to cache-bust requests on 2946 websites.}
    \label{tab:cache_busting}
    \centering
    \begin{tabular}{lrr}
    \toprule

    Cache-busting technique & Cache busted \\

    \midrule

    Query string                        & 2112 (60.4\%) \\
    \texttt{Origin} header              & 817 (23.4\%) \\
    \texttt{User-Agent} header          & 78 { } (2.2\%) \\
    \texttt{X-Forwarded-Host} header    & 327 { } (9.4\%) \\
    \texttt{X-Forwarded-Scheme} header  & 329 { } (9.4\%) \\
    \texttt{X-Method-Override} header   & 338 { } (9.7\%) \\
    Headers in \texttt{Vary} header     & 616 (17.6\%) \\
    
    \midrule

    All techniques combined & 2946 (84.3\%) \\

    \bottomrule
    \end{tabular}
\end{table}

Table~\ref{tab:cache_busting} presents the effectiveness of different cache-busting techniques, based on our experiment on the Tranco Top 10k. We identified a cached response on 3494 websites. Collectively, with the techniques that we employed, we were able to cache-bust requests on 2946 websites. On the 3128 websites that included the \texttt{Vary} header in their responses, introducing modifications in the headers included in it effectively cache-busted the requests only on 616 websites. Therefore, 2512 websites include some header names in the \texttt{Vary} response header but do not configure their web cache(s) to include them in the cache key. In our experiments, we use all techniques combined to maximise the likelihood of effectively cache-busting the request.

\subsubsection*{HTTP Multiplexing}

To send two requests in the same TCP packet, we use the same methodology exploited by Van Goethem et al. to perform \textit{timeless timing attacks} in~\cite{tta}, sending two \texttt{HEADERS} frames containing the two HTTP/2 requests in a single packet. Sending the two requests within a single TCP packet eliminates the effect of network latency and jitter, enabling timing measurements that are not influenced by them. Our methodology can also be implemented using HTTP/3 multiplexing, which enables simultaneously sending multiple requests over a single connection.

\subsection{Reading the Timings}
\label{sec:methodology:phase_2}

In the second phase, we read the time measurements collected in the first phase to infer whether the responses to the sent requests are coming from a web cache or if they originated from the backend server. To do this, we employ a statistical test.

\subsubsection*{Statistical Test}

In this step, we use a \texttt{t-test} to determine whether there is a statistically significant difference between the means of the measurements of the \randomized and \fixed group of paired requests. If the difference is significant, we conclude that there is a cache in the path from the client to the origin server (we set the threshold of the p-value to 0.01). Otherwise, we conclude that there is no cache. We use the t-test as a classifier where, depending on a set threshold, we can classify requests as cached or not.
Before performing the statistical test, to enhance its accuracy, we employ two heuristics to pre-process the data. First, we remove the time measurements with outlier values~\footnote{In particular, we compute the data's average and standard deviation and the absolute difference between the time difference and the average. If the absolute difference is lower than the standard deviation multiplied by a factor of 2, the data point is considered an outlier and removed. The factor was selected manually to remove the highest number of outliers while not losing too many data points.} to prevent delayed packets (that might be due to network congestion, server overload and other unpredictable conditions) from negatively influencing our classification. Then, we multiply the negative values in the \fixed group measurements by a factor of 5 when the group's mean is negative.
We expect negative values when a response is cached because the request with a fixed cache buster is placed as second, and it will arrive before the response to the first request.
We do this to amplify negative values, making them more significant and easing the statistical test classification. The factor value is based on our preliminary experiments and observations. In the \fixed group, when a cache is present, we expect negative timing measurements with low standard deviation.

\section{Experiments}
\label{sec:experiments}

\begin{table*}[t]
    \caption{On the left, a sample of the time measurements of a timing attack against a website that presents a web cache; on the right, a sample where the responses are not cached. We can observe that, where the responses in the \fixed group are cache, while the timings for the \randomized group are extremely variable, the timings for the \fixed group are consistent (i.e., negative with a low standard deviation). When instead all the responses are coming from the origin server, we see that both the \randomized and \fixed group's time measurements are inconsistent. Negative timings mean the response to the second request in the pair arrived first. \textit{CHH} refers to \textit{cache header heuristics} algorithm from~\cite{mirheidari2022web}. For brevity, this example only shows 5 time measurements for each group of paired requests, while in all our experiments we sent 10. The two samples are from different websites.}
    \label{tab:examples}
    \centering
    \setlength\tabcolsep{3pt}
    \begin{tabular}{lrcc@{\hspace{0.5cm}}l@{\hspace{0.5cm}}lrcc}

    \toprule

    \multicolumn{4}{c}{\textbf{Time measurements with cached responses}} & & \multicolumn{4}{c}{\textbf{Time measurements with \textbf{no} cached responses}} \\

    \noalign{\bigskip}

    Group & Time diff. (ms) &  Cache Status 1  &  Cache Status 2  & & Group & Time diff. (ms) &  Cache Status 1  &  Cache Status 2 \\

    \cline{1-4} \cline{6-9} \noalign{\smallskip}

                 & -60.09 &       MISS       &       MISS         &  &             &        34.37   &       MISS       &       MISS       \\
                 &  62.42 &       MISS       &       MISS         &  &             &        97.29   &       MISS       &       MISS       \\
    Randomized   & -58.35 &       MISS       &       MISS         &  & Randomized  &        -486.03 &       MISS       &       MISS       \\
                 &  67.32 &       MISS       &       MISS         &  &             &        132.2   &       MISS       &       MISS       \\
                 & -77.45 &       MISS       &       MISS         &  &             &        -325.18 &       MISS       &       MISS       \\

    \cline{1-4} \cline{6-9} \noalign{\smallskip}

            & -600.95 &       MISS       &       HIT         &  &        &       -169.52 &       MISS       &       MISS       \\
            & -504.63 &       MISS       &       HIT         &  &        &       12.2    &       MISS       &       MISS       \\
    Fixed   & -591.15 &       MISS       &       HIT         &  & Fixed  &       -409.99 &       MISS       &       MISS       \\
            & -516.49 &       MISS       &       HIT         &  &        &       -31.29  &       MISS       &       MISS       \\
            & -536.35 &       MISS       &       HIT         &  &        &       217.21  &       MISS       &       MISS       \\

    \cline{1-4} \cline{6-9} \noalign{\bigskip}

    Method &  Cache Status  &  {} & & & Method &  Cache Status  &  {}  \\

    \cline{1-4} \cline{6-9} \noalign{\smallskip}

    CHH               &     \textit{Cache}      &                & & &   CHH               &     \textit{No Cache}      &               \\
    Statistical test  &     \textit{Cache}      &      {}   & & &  Statistical test  &     \textit{No Cache}      &      {}  \\

    \bottomrule

    \end{tabular}
\end{table*}

In this section, we describe the experiments we performed to test our methodology for detecting web caches and report the results. First, we carried out a preliminary experiment on websites in the Tranco Top 10k~\cite{tranco}~\footnote{In all our experiments, we use the list generated on 29 January 2024, available at https://tranco-list.eu/list/QGN64.} that report the cache status of their responses in the headers to validate our hypothesis that timing analysis can be used to detect web caches and measure the accuracy of this technique.

Next, we performed a large-scale measurement on the Tranco Top 50k to measure the prevalence of hidden web caches. For the websites that advertise the cache status of responses in the headers, we also check if the classification matches the reported status to better estimate the accuracy of our detection technique.

In both experiments, our crawler runs without authentication, meaning that it can only visit and test the web pages that are publicly accessible. Our crawler is developed in Python and uses the \texttt{requests} library to perform the HTTP requests. To avoid our requests being blocked by the websites, we use a real browser user agent and we limit the number of performed requests per second. We only tested websites that support HTTP/2 in our experiments because of the availability of off-the-shelf libraries and because it is more widely adopted compared to HTTP/3; we leave the task of implementing an HTTP/3 version of our tool for future work.

Table~\ref{tab:examples} presents an example of timing measurements in a scenario where the second response of the \fixed group is coming from a cache (on the right) and one where all the responses are coming from the origin server. We can observe that, when one of the two responses of two paired requests is cached, the time measurements are all negative and close to each other with a low standard deviation. 

\subsection{Preliminary Experiment}
\label{sec:experiments:preliminary_experiment}

\begin{table}[t]
    \caption{The results of the preliminary experiment over the Tranco Top 10k. *The percentages of ``Reachable'', ``Tested'', ``Analysed'' and ``Discarded'' are calculated over the 10k websites from the Tranco list. The percentages of ``Correct classification'' and ``Wrong classification'' are calculated over the number of correctly analysed sites (1.946). 2.621 domain names could not be reached because they timed out, did not listen for HTTP requests or had other errors.}
    \label{tab:preliminary_experiment}
    \centering
    \begin{tabular}{lrr}
    \toprule

     & Number of sites & Percentage* \\

    \midrule

    Reachable            &    7.379 & 73.8\%      \\
    Tested               &    2.289 & 22.9\%      \\
    Discarded            &     343 & 17.6\%      \\
    Analysed             &    1.946 & 19.5\%      \\

    \midrule
    
    \multicolumn{1}{r}{\textit{Correct classification}} &    1.743 & 89.6\%       \\
    \multicolumn{1}{r}{\textit{Wrong classification}}   &     203 & 10.4\%       \\

    \bottomrule
    \end{tabular}
\end{table}

This preliminary experiment aims to understand if it is possible to infer the presence of a web cache by timing the responses. Moreover, we measure the accuracy of the t-test in classifying whether a cache is present or not. To do this, this experiment only targets websites that present cache status headers in their responses and uses it as a ground truth for the cache status of the responses. If during the crawling phase we do not identify a request that gets cached, we issue a request to a non-existent path to obtain a \textit{404 Not Found} response (frequently cached). We do this because, in this experiment, we are not interested in the content of the cached response, but only in the response being cached.

During this experiment, we observed multiple websites issuing most likely wrong or untruthful cache status headers, only reporting cache misses, which lowered the accuracy value of the t-test. We investigated multiple cases where, based on the time measurements, a cache is almost certainly present to understand why the cache headers do not report the cache HITs. We observed that this is mainly due to CDNs only reporting cache MISSes in the responses to paired requests, even when the responses are cached. Since CDNs are black boxes for us, we cannot provide possible explanations for this behaviour.
To better estimate the real accuracy of the t-test in classifying cached responses based on time measurements, we then selected 100 websites for which the cache headers report only MISSes, while the t-test reports the presence of a cache, and manually verified if a cache is indeed present by sending requests in single packets and inspecting their cache status headers.

In this experiment, we crawl the websites from the Tranco Top 10k starting from their homepage and, for each URL, we check if it reports the cache status headers. If it does, we collect the time measurements. Recall that we send two batches of paired requests, one where both paired requests are served by the origin, and the second where one of the two paired requests is instead served by the cache. Based on our initial experiments and observations, we set the number of request pairs \textbf{n} to 10 for each batch. Experimentally, we observed that a smaller number of request pairs deteriorated the accuracy of our tool, while more requests did not provide significant improvements.

In the first batch, we expect both responses of paired requests to be cache MISS, i.e., they come from the origin, and in the second we expect one cache MISS and one cache HIT, i.e., one response from the origin and one from the cache. We discard from our data all the measurements with more than one wrong cache status in any of the two batches since we observed that a single error in the cache statuses does not compromise the overall quality of the classification. This number is a good compromise between the stress imposed on the tested websites and the accuracy of our classification. To limit the overhead caused by our tests on the targeted websites, we also limit our crawler to test at most 10 URLs on at most 10 FQDNs for every root domain in the Tranco list. It is important to note that we stop our tests once we find one request that gets cached.

\subsubsection{Results}

Table~\ref{tab:preliminary_experiment} presents the results of our preliminary experiment. Of the 10k domain names in the Tranco Top 10k list, only 7.379 were reachable; the remaining 2.621 domain names timed out, did not respond to HTTP requests or redirected to another domain included in the list. Of the reachable websites, 2.289 present cache status headers and support HTTP/2 and were therefore tested. We then discarded 343 websites from our data due to responses with the wrong cache status (i.e., a cached response where we expect a response coming from the origin, or vice-versa), resulting in 1.946 correctly analysed websites. On these, our methodology correctly identified the cache status of the responses on 1.743 websites (89.6\%) and failed on 203 websites. However, as we mentioned previously, looking at the time measurements we hypothesise that a high percentage of these wrongly labelled measurements are due to the caches advertising a wrong or untruthful cache status, rather than our methodology failing to correctly identify it. For this reason, we selected 100 websites and manually analysed them to validate our methodology's conclusion and found that 82 wrongly labelled measurements were due to wrong cache statuses being advertised by a web cache, and were therefore false negatives. Based on this data, we estimate that the accuracy of our technique in identifying the cache status of responses only based on timing analysis is higher compared to the one reported, which should be considered as a lower bound.
This experiment answers our first research question~\ref{goal:one}, showing that timing analysis is highly effective in detecting cached responses and distinguishing them from responses originating from the backend server.

\subsection{Large-scale Experiment}
\label{sec:experiments:large-scale_experiment}

In this experiment, we do not filter the websites based on the presence of cache status headers in the responses, but we employ our methodology on all the websites in the Tranco Top 50k. The goal of this experiment is to discover \textit{hidden caches} that do not advertise the cache status headers in their responses. Similarly to the preliminary experiment, we set the number of request pairs \textbf{n} to 10 for each group and limited our crawler to test at most 10 URLs on at most 10 FQDNs for every root domain in the Tranco list. For each URL, we collect the timing measurements and use the statistical test to infer the cache status of the response. We also check if the classification matches the cache status reported in the headers if they are present. It must be noted that, due to the possible absence of cache status headers, we have no way of checking if the cache-busting techniques that we apply to the requests are effective or not. For this reason, it is possible that in some cases it will appear as though there is no cache, while in reality a cache is present and cannot be detected since no response is coming from the origin server (i.e., there is no timing difference to observe between the \fixed and \randomized groups).

In this experiment, we use the t-test as we did in the preliminary experiment and we apply the same pre-processing techniques to the time measurements.

\subsubsection{Results}

\begin{table}[t]
    \caption{The results of the large-scale experiment over the Tranco Top 50k. *The percentages of ``Reachable'' and ``Tested'' calculated over the 50k websites from the Tranco list. All the other percentages are calculated over the number of correctly analysed sites (28.243). 10.841 domain names could not be reached because they timed out, did not listen for HTTP requests or had other errors.}
    \label{tab:experiment}
    \centering
    \begin{tabular}{lrr}
    \toprule

     & Number of sites & Percentage*  \\

    \midrule

    Reachable            &    39.159 & 78.3\%      \\
    Tested               &    28.243 & 56.5\%      \\

    \midrule

    Present cache status headers                        & 10.543 & 37.3\%    \\
    \multicolumn{1}{r}{\textit{Correct classification}} &  7.280 & 25.8\%     \\
    \multicolumn{1}{r}{\textit{Wrong classification}}   &  3.263 & 11.6\%     \\

    \midrule

    No cache status headers                 & 17.700 & 62.7\%       \\
    \multicolumn{1}{r}{\textit{Cache}}      &  1.627 &  5.8\%       \\
    \multicolumn{1}{r}{\textit{No cache}}   & 16.073 & 56.9\%       \\

    \bottomrule
    \end{tabular}
\end{table}

Table~\ref{tab:experiment} presents the results of our large-scale measurement experiment on the Tranco Top 50k. Of these 50k domain names, 10.841 could not be reached due to timeouts, redirects, and the absence of an HTTP server listening. We successfully visited 39.159 domains, 28.243 of which supported HTTP/2 and could therefore be tested with our methodology.

Of the 28.243 correctly tested websites, 10.543 advertise cache status headers in their responses, which we compared against the classification of our methodology. Using the cache status headers as a ground truth (that, as we saw in our preliminary experiment, is only the best compromise available), 7.280 were correctly classified by our methodology (69.1\% over the 10.543 websites that present cache status headers), while for the remaining 3.263 websites (30.9\%), our methodology gave an incorrect classification.
We hypothesize that the accuracy in this experiment is lower due to the lower popularity of the websites tested, which might use less efficient cache technologies and make less use of CDNs. For example, less popular websites might employ web caches placed on the same machine as the origin server. This results in smaller differences in the time measurements for cached responses and responses coming from the origin server. 
In this experiment, due to the larger sample size tested, we did not perform a manual validation of the results.

17.700 websites that we correctly tested did not advertise the cache status of their responses in the response headers (or used custom names and values for their headers that are not covered by the \textit{cache header heuristics}). Of these, our methodology classified 1.627 websites as having a hidden web cache, and the remaining 16.073 as not having a web cache.
This answers our second research question~\ref{goal:two}, estimating the prevalence of hidden caches at \textbf{5.8\%}.

\section{Vulnerabilities Detection}
\label{sec:vulnerabilities_detection}

Now that we have a methodology that detects the presence of hidden web caches, we use it to detect vulnerabilities that would otherwise be extremely difficult to spot. Specifically, we use it to detect Web Cache Deception (WCD) vulnerabilities. Web cache vulnerabilities are generally extremely complicated to detect automatically, mainly because these vulnerabilities exist when the cache interacts with different components of an architecture in a complex system (e.g., an origin server, another web cache, or a proxy). For WCD, instead, previous studies have presented effective automated detection methodologies~\cite{mirheidari2020cached, mirheidari2022web}. In this section, we describe how we tested the identified hidden caches for WCD vulnerabilities and present the results of our analysis.

\subsection{Detection Methodology}
Algorithm~\ref{alg:methodology} presents a simplified pseudo-code for our detection methodology.
We crawl each site to test and, for each URL that we visit, similarly to Mirheidari et at.~\cite{mirheidari2022web}, we test whether the response includes dynamic content or if the page is static. We do this by performing a simple string comparison of the responses. If the response is dynamic, we generate two attack URLs (lines 1-2), i.e., we modify the URL, including a path confusion payload and a WCD payload. A WCD payload comprises a non-existent file name and a static file extension (we use \texttt{.css}). Again, we check whether the response to the attack URLs is dynamic (lines 3-5). We perform this check because a WCD attack aims to induce a web cache into storing dynamic content that could contain sensitive data; a static file that is the same for all visitors is unlikely to include sensitive information.
If the response to the attack URL is dynamic, we proceed with the timing analysis. The idea behind this methodology is to detect whether a response that includes dynamic content is cached and, therefore, if the website is vulnerable to WCD. Note that this definition of Web Cache Deception is wide and does not consider what data is mistakenly cached; instead, it is based on the idea that a web cache should not publicly cache dynamic pages.

Next, we send two groups of paired requests as follows:

\begin{enumerate}
    \item \textbf{\texttt{n}} pairs of requests in a single packet to the base URL (i.e., the URL without added payloads), where both requests have random cache busters, i.e., the responses should always be served by the origin server (line 7-8).
    \item \textbf{\texttt{n}} pairs of requests in a single packet where the first request has a random cache buster and the second request is to a single fixed attack URL (generated at line 10), i.e., the first response should be served by the origin server, the second by the cache \textit{if the website is vulnerable to WCD} (lines 12-13).
\end{enumerate}

We observe the order of arrival and measure the time elapsed between receiving the responses of paired requests and, using the t-test, we check if there is a significant difference in the timings between the two groups of paired requests (lines 15-16). If the difference is significant, we conclude that the response to the WCD-payloaded request is cached and, therefore, the website is vulnerable to Web Cache Deception.

\begin{algorithm}
  \small
  \caption{Simplified pseudo-code for our WCD detection methodology based on timing analysis. \textit{$\alpha$} is the significance level for the statistical test.}
  \label{alg:methodology}
  \SetAlgoLined
  \SetKwInOut{Input}{input}
  \Input{URL}
  \SetKwFor{RepTimes}{repeat}{times}{end}
  
$attackURL1 = generateAttackURL(URL)$\;
$attackURL2 = generateAttackURL(URL)$\;

$result1 \gets get(attackURL1)$\;
$result2 \gets get(attackURL2)$\;

\If{$result1 \neq result2$}{
    $timings1 = []$\;

    \RepTimes{n}{
        $timings1.append(timingAnalysis(\newline cacheBust(URL), cacheBust(URL)))$\;
    }

    $attackURL = generateAttackURL(URL)$\;

    $timings2 = []$\;
    \RepTimes{n}{
        $timings2.append(timingAnalysis(\newline cacheBust(URL), attackURL))$\;
    }

    \If{$t\_test(timings1, timings2) \leq \alpha$}{
        \Return{WCD detected}
    }
}
\end{algorithm}

\subsection{Experiment}
We tested all 1.627 sites where we detected the presence of a hidden cache during our large-scale analysis of the Tranco Top 50k. Similarly to the previous experiments, we set the number of request pairs \textbf{n} to 10, and limit our crawler to visit at most 10 URLs on at most 10 FQDNs for every website that we test to limit the load imposed by our analysis on the servers. We use the following WCD payloads in our tests, selected based on our previous experience and experiments:

\begin{itemize}
    \item Path parameter: \texttt{/}
    \item Encoded question mark: \texttt{\%3F}
    \item Encoded semicolon: \texttt{\%3B}
\end{itemize}

\subsection{Results and Case Studies}
Of the 1.627 sites that our methodology detected as presenting a hidden cache, we detected that 1.020 (62.7\%) are caching dynamic content. As black-box testers, we can't understand whether these websites are doing that deliberately, or if that indicates a Web Cache Deception vulnerability. Testing all the affected websites for WCD vulnerabilities would be an extremely onerous manual task, and it is out of the scope of our research.

To validate our findings, we manually analysed a subset of these websites, limiting our tests to 35 randomly sampled sites among the ones caching dynamic content. In this analysis, our goal is to understand whether the identified vulnerabilities can be exploited to steal victims' sensitive information. Even if a site is caching dynamic information, this does not directly imply that the site is leaking private data, and the dynamic parts of web pages could be harmless content, such as timestamps and random error codes.

We registered test accounts on the sites that did not require personal information during the registration phase (e.g., a valid phone number, or a payment card), excluding the others. We successfully created a test account on 19 websites.
Next, we checked how many of them cache dynamic content of pages also when the visitor is authenticated, resulting in 15 websites.
Finally, we manually tested these websites to check if they leak sensitive information of authenticated users, finding that 5 websites are vulnerable to Web Cache Deception and leak private data.

Following we briefly describe the consequences of the vulnerabilities discovered during the manual analysis.

\paragraph{Case Studies}
Our tests uncovered three large e-commerce websites vulnerable to WCD attacks. All of them leak personal information of the target victims, such as their email, geographical location and their shopping cart. Moreover, we discovered a large micro-blogging website vulnerable to WCD that leaked the emails of the target victims. It is important to specify that these vulnerabilities could not have been discovered using the state-of-the-art WCD detection techniques, due to the lack of cache status headers in the sites' HTTP responses. This highlights that our novel methodology is crucial for identifying well-hidden vulnerabilities that would otherwise be impossible to spot and detect.

\section{Discussion}

All state-of-the-art techniques to detect whether an HTTP is cached or not rely on cache status headers. These techniques are ineffective when web caches do not communicate the cache status of responses, when the headers they use have custom names and values, or when the cache status communicated is wrong. We developed a novel methodology to overcome this limitation that can distinguish between cached and non-cached responses using timing analysis, without relying on the response headers.

We first performed a preliminary experiment on websites that present cache status headers to investigate whether timing analysis is a viable technique to distinguish between cached and non-cached responses. Comparing the classification of our tool with the cache status communicated in the response headers, we estimated an accuracy of 89.6\%. Moreover, we manually verified the cache status headers on 100 randomly selected websites where our methodology gave an incorrect classification, finding that in 82\% of the cases it was due to the website communicating a wrong cache, rather than a misclassification of our methodology. This hints to us that the real accuracy of our tool is higher compared to the one we measured.
We answer our first research question~\ref{goal:one} affirmatively, concluding that timing analysis is highly accurate in detecting cached responses.

We then used our novel methodology and tool to investigate the prevalence of \textit{hidden caches} on the Tranco Top 50k. We use the term \textit{hidden caches} to refer to those web caches detected by our tool that do not communicate the cache status of responses in the headers. On 1.627 websites that did not present cache status headers in their responses, we detected a cached response, hinting at the presence of a hidden web cache. We can therefore estimate that hidden caches are present on 5.8\% of websites that support HTTP/2 in the Tranco Top 50k, answering our second research question~\ref{goal:two}.

Finally, we employed our novel methodology to detect Web Cache Deception vulnerabilities in the hidden caches identified during our large-scale experiment on the Tranco Top 50k, finding that 1.020 websites out of 1.627 (62.7\%) cache dynamic responses. Caching dynamic responses is not necessarily an indication of a vulnerability, but it certainly is an unusual behaviour that, in specific circumstances, might lead to the leakage of sensitive information of the websites' visitors. We manually investigated 35 randomly sampled websites that cache dynamic data and identified 5 websites vulnerable to WCD attacks that could be exploited to leak sensitive information about victims. We find that hidden caches are affected by common cache vulnerabilities and a vast majority of them cache content that they should not, answering our third research question~\ref{goal:three}.

The methodology that we present can be useful for website owners and security researchers to identify well-hidden vulnerabilities that would otherwise be impossible to spot. Our results confirm that cache vulnerabilities are highly prevalent on the web, and require major efforts by website operators not to expose internet users to attacks that might violate their privacy and security.

\subsection{Ethical Considerations}
During our experiments, we minimized the impact  of our tests on the load of the targeted servers by limiting the number of requests performed as much as possible and by slowing our requests to no more than two paired requests every half a second. Specifically, during the crawling phase, we limit our tool to visit at most 10 pages on at most 10 FQDNs, for a maximum of 100 requested web pages. In reality, as most of our tests were performed on a single page for each domain in the Tranco list, we stop the crawling phase once we identify a candidate page where we can successfully perform our tests. In both experiments, our timing analysis comprised two runs of 10 paired requests each, resulting in 40 requests sent. We argue that this number is a good compromise between the accuracy of the timing analysis and the excess load introduced on the servers by our tests. When testing for web cache vulnerabilities, we never attacked real users of the target websites but always used test accounts controlled by us. We never leaked, or tried to leak, any personally identifiable information of real internet users and we never injected any malicious payload in the tested web caches.
Finally, we are currently in the process of responsibly disclosing all the identified Web Cache Deception vulnerabilities that lead to the possible leakage of victims' sensitive information to the impacted parties through their coordinated vulnerability disclosure channels.

\section{Conclusions}

In this paper, we presented a novel methodology for detecting cached responses using timing analysis. This methodology overcomes the limitations of previous approaches that rely on cache status headers, which are not standardized and can be missing or unreliable. Our method applies to any web server that supports HTTP/2 or HTTP/3, regardless of its cache disclosure practices.
We developed a timing analysis-based methodology that achieves an estimated accuracy of 89.6\% in differentiating between cached and non-cached responses. We identified an uncommon behaviour where certain web caches only report cache MISSes for paired requests, even if one response is a cache HIT, highlighting the limitations of solely relying on cache status headers to detect caching. Using our methodology, we estimated that 5.8\% of websites within the Tranco Top 50k that support HTTP/2 employ hidden caches that do not advertise their presence through cache status headers. We leveraged our timing analysis methodology to detect Web Cache Deception (WCD) vulnerabilities in a black-box manner, discovering that 1.020 of the identified hidden caches were susceptible to WCD vulnerabilities, potentially leading to leakage of victims' sensitive data.
Our findings demonstrate the effectiveness of our novel timing analysis methodology for cache detection and WCD vulnerability identification. This methodology provides a valuable tool for security researchers and website operators to assess caching behaviours and identify potential security risks.

\begin{acks}
We acknowledge the support of the MUR PNRR project PE SERICS - SecCO (PE00000014) CUP D33C22001300002 funded by the European Union under NextGenerationEU. Views and opinions expressed are however those of the author(s) only and do not necessarily reflect those of the European Union or European Commission. Neither the European Union nor the granting authority can be held responsible for them.
\end{acks}

\bibliographystyle{ACM-Reference-Format}
\balance
\bibliography{paper}

\end{document}